\documentstyle[12pt]{article}

\newcommand{\sect}[1]{\setcounter{equation}{0}\section{#1}\indent}

\newcommand{\EQ}{\begin{equation}}
\newcommand{\EN}{\end{equation}}
\newcommand{\bea}{\begin{eqnarray}}
\newcommand{\ena}{\end{eqnarray}}
\newcommand{\vs}[1]{\vspace{#1 mm}}
\renewcommand{\a}{\alpha}
\renewcommand{\b}{\beta}
\renewcommand{\c}{\gamma}
\renewcommand{\d}{\delta}

\newcommand{\shalf}{\frac{1}{2}}
\newcommand{\pa}{\partial}
\renewcommand{\t}{\theta}
\newcommand{\tb}{{\bar \theta}}
\newcommand{\dz}{\frac{dz}{2\pi i}}
\newcommand{\dZ}{\frac{dzd^2\t}{2\pi i}}
\newcommand{\dZc}{\frac{dzd \bar \t}{2\pi i}}
\newcommand{\dZa}{\frac{dzd\t}{2\pi i}}
\newcommand{\nn}{\nonumber\\}

\begin{document}

\topmargin 0pt
\oddsidemargin 5mm

\newcommand{\NP}[1]{Nucl.\ Phys.\ {\bf #1}}
\newcommand{\AP}[1]{Ann.\ Phys.\ {\bf #1}}
\newcommand{\PL}[1]{Phys.\ Lett.\ {\bf #1}}
\newcommand{\NC}[1]{Nuovo Cimento {\bf #1}}
\newcommand{\CMP}[1]{Comm.\ Math.\ Phys.\ {\bf #1}}
\newcommand{\PR}[1]{Phys.\ Rev.\ {\bf #1}}
\newcommand{\PRL}[1]{Phys.\ Rev.\ Lett.\ {\bf #1}}
\newcommand{\PTP}[1]{Prog.\ Theor.\ Phys.\ {\bf #1}}
\newcommand{\PTPS}[1]{Prog.\ Theor.\ Phys.\ Suppl.\ {\bf #1}}
\newcommand{\MPL}[1]{Mod.\ Phys.\ Lett.\ {\bf #1}}
\newcommand{\IJMP}[1]{Int.\ Jour.\ Mod.\ Phys.\ {\bf #1}}
\newcommand{\JP}[1]{Jour.\ Phys.\ {\bf #1}}

\begin{titlepage}

\setcounter{page}{0}
\begin{flushright}
 NBI-HE-94-10
\end{flushright}

\vspace{1cm}
\begin{center}
{\Large THE BRST OPERATOR FOR THE LARGE $N=4$ SUPERCONFORMAL ALGEBRA }
\vspace{1.5cm}

{\large Fiorenzo Bastianelli$^{a,}$\footnote{e-mail address:
fiorenzo@nbivax.nbi.dk} and Nobuyoshi Ohta$^{b,}$\footnote{e-mail address:
ohta@nbivax.nbi.dk, ohta@fuji.wani.osaka-u.ac.jp}$^,$\footnote{Permanent
address: Department of Physics, Osaka University, Toyonaka, Osaka 560,
Japan.}} \\
\vspace{1cm}
$^a${\em The Niels Bohr Institute, Blegdamsvej 17, DK-2100 Copenhagen \O,
    Denmark}\\
\vspace{.5cm}
$^b${\em NORDITA, Blegdamsvej 17, DK-2100 Copenhagen \O, Denmark}\\

\end{center}
\vspace{15mm}
\centerline{{\bf{Abstract}}}
\vspace{.5cm}

We review the detailed structure of the large $N=4$ superconformal
algebra, and construct its BRST operator which constitutes the main
object for analyzing $N=4$ strings. We then derive
the general condition for the nilpotency of the BRST operator and
show that there exists a line of critical $N=4$ string theories.

\end{titlepage}
\newpage
\renewcommand{\thefootnote}{\arabic{footnote}}
\setcounter{footnote}{0}

\sect{Introduction}
\indent
It has long been known that bosonic strings and superstrings share
many common features. For example, it is often observed that
any calculation that can be done in bosonic strings can also be
repeated in superstrings with just a few additional technical details.
It is tempting then to speculate that all these string theories are
deeply related and could be considered as different phases of a single
unified theory. In fact, it has been shown that string theories with
$N=0,1$ supersymmetries can be embedded into those with $N=1,2$,
respectively~[1-5]. It has also been shown that $N=2$ superstrings can
be regarded as  spontaneously broken phases of $N=4$ strings~\cite{BOP},
and, more recently, that there exists a hierarchy of embeddings in which
the $N$ string is embedded into the $N+1$ string~\cite{BOP2}.
These $N$ strings are the ones based on the superconformal algebras
(SCA) found by Ademollo et al.~\cite{N4S}.
Out of these algebras the most intriguing is the $N=4$, which admits
two independent central extensions~\cite{STV} (and can be reduced to the
so-called small $N=4$, admitting only one central extension and
possessing an $SU(2)$ Kac-Moody algebra).
For $N>4$ there are no possible central extensions and
the corresponding BRST charge is automatically nilpotent.
It is the purpose of this paper to analyze instead the BRST operator
for the large $N=4$ SCA (i.e. the one with two central extensions),
which is the key object to examine
the physical states and to construct BRST invariant vertices for the
$N=4$ superstring. This is, in fact, the $N=4$ superstring
that enters the hierarchy constructed in ref.~\cite{BOP2}.
Therefore, we are going to summarize the detailed structure of
the large $N=4$ SCA in terms of components as well as in
$N=2$ superfields.
We present also the explicit connection between the two descriptions.
We then construct the BRST operator for the corresponding string
theory and obtain the condition for its
nilpotency. This analysis shows the existence of a one-parameter family
of critical $N=4$ string theories. As an aside, we present a
useful technique which allows us to immediately write down the
BRST operator once the operator algebra is given.
We believe that our results will be useful in further studies of
the $N=4$ string theory.

The paper is organized as follows. In sect. 2, we review the
component description of the large $N=4$ SCA and derive its BRST
operator. Our method of derivation is explained in appendix A.
We then present in sect. 3 the equivalent results in terms of
$N=2$ superfields. The nilpotency condition for the BRST operator
is examined in sect. 4. Finally, sect. 5 reviews how to recognize
the small $N=4$ SCA as a subalgebra of the large $N=4$ SCA
using $N=2$ superfields.
This may be useful in trying to construct
an embedding of the small $N=4$ string into the large one.
The ghost generators which appear in the BRST charges are collected
in appendix B.

\sect{Large $N=4$ SCA and its BRST operator in components}
\indent
In this section, we first summarize the large $N=4$
SCA and construct its BRST operator.
The algebra is well known~\cite{N4S,STV,STVS}, but we try
to make the group structure clearer. This also serves to establish
our notation.

The large $N=4$ SCA consists of the
energy-momentum tensor $T$, four supersymmetry generators $G_a,\ (a=(1,1),
(2,1),(1,2),(2,2))$, two commuting sets of $SU(2)$
currents $J^{A,i}$,
($A= (+,-)$ distinguishes between the two sets of currents and
$ i= (+,-,3)$  is the index for the $SU(2)$ generators in the Cartan basis),
four fermionic generators $F_a$
and one $U(1)$ current $J$. The fermionic generators $G_a$ and $F_a$ are
doublets with respect to the two $SU(2)$ currents and the first and
second entries in their suffices refer to these two $SU(2)$.
The operator products (OPEs) for the $N=4$ algebra in these components reads
\bea
T(z) T(w) &\sim&\
\frac{\frac{1}{2}c}{(z-w)^4}
+ \frac{2T(w)}{(z-w)^2} + \frac{\pa T(w)}{(z-w)}\ ,\nn
T(z) {\cal O}(w) &\sim&\
\frac{h_{\cal O} {\cal O}(w)}{(z-w)^2}+\frac{\pa {\cal O}(w)}{(z-w)}\ ,\nn
J^{\pm,i}(z) J^{\pm,j}(w) &\sim&\
\frac{\frac{1}{2} k^\pm g^{ij}}{(z-w)^2}
+ \frac{f^{ij}{}_k J^{\pm,k}(w)}{(z-w)}\ ,\nn
J^{\pm,i}(z) G_a(w) &\sim&\
{\mp 2 {k^\pm \over k} R^{\pm,i}{}_a{}^b F_b(z) \over (z-w)^2} +
{R^{\pm,i}{}_a{}^b G_b(w)\over (z-w)}\ ,\nn
J^{\pm,i}(z) F_a(w) &\sim&\
{ R^{\pm,i}{}_a{}^b F_b(w)\over (z-w)}\ ,\nn
G_a(z) G_b(w) &\sim&\
{{2 \over 3}c \eta_{ab} \over (z-w)^3} + {2 M_{ab}(w) \over (z-w)^2} +
{2 \eta_{ab} T(w) + \pa M_{ab}(w) \over (z-w)}\ ,\nn
F_a(z) G_b(w) &\sim&\ {2 R^{-,i}_{ab} J^-_i(w)-
2 R^{+,i}_{ab} J^+_i(w) + \eta_{ab} J(w) \over (z-w)}\ ,\nn
F_a(z) F_b(w) &\sim&\ {- {1\over 2}k\eta_{ab}\over (z-w)}\ ,\nn
J(z) G_a(w) &\sim&\ { F_a(w)\over (z-w)^2}\ ,\nn
J(z) J(w)\ &\sim&\ { -{1\over 2}k \over (z-w)^2}\ ,
\ena
where ${\cal O}$ stands for the generators $G, J, F$ with their dimensions
given by $h_{\cal O}$:
\bea
h_{\cal O} &=&
\left\{
\begin{array}{lll}
\frac{3}{2} & {\rm for} & G_a, \\
1 & {\rm for} & J^{A,i},J, \\
\frac{1}{2} & {\rm for} & F_a,
\end{array} \right. \\
M_{ab} &=&  4 {k^- \over k}R^{+,i}_{ab} J^+_i
+4 {k^+ \over k}R^{-,i}_{ab} J^-_i \ , \\
k &=& k^+ + k^-\ , \ \  c = \frac{6k^+ k^-}{k} \ .
\ena
In the Cartan basis we have the following components for the $SU(2)$
Killing metric and structure constants
\EQ
g^{+-} = 2 \ , \ g^{33} = 1\ ; \ \ f^{+-}{}_3 = 2 \ , \
f^{3\pm}{}_\pm = \pm 1\ ,
\EN
while other components not related by symmetry are zero.
Using the double index notation $ a=(\a,\bar\a)$,
the $SU(2)$ representation matrices $R^{\pm,i}{}_a{}^b$
can be written as follows
\EQ
 R^{+,i}{}_{(\a,\bar \a)}{}^{(\b, \bar \b)} =
\left\{ \begin{array}{ll}
\frac{1}{2} \bar \sigma^i{}^\b{}_\a & \mbox{if $ \bar\a = \bar\b = 1$} \\
\frac{1}{2}  \sigma^i{}^\b{}_\a & \mbox{if $\bar \a =\bar \b =2$} \\
0 & \mbox{otherwise}
\end{array} \right.\ , \ \
 R^{-,i}{}_{(\a,\bar \a)}{}^{(\b, \bar \b)} =
\left\{ \begin{array}{ll}
\frac{1}{2} \bar \sigma^i{}^{\bar \b}{}_{\bar \a}
 & \mbox{if $ \a = \b = 1$} \\
\frac{1}{2}  \sigma^i{}^{\bar \b}{}_{\bar \a}
 & \mbox{if $\a = \b =2$} \\
0 & \mbox{otherwise}
\end{array}
\right.\ ,
\EN
where $\sigma^i = (\sigma^3, \sigma^+ ,
\sigma^-)$ are the Pauli matrices in the Cartan basis
and $\bar \sigma^i  = ( \sigma^3, -\sigma^+,-\sigma^-)$.
Finally, the invariant tensor $\eta_{ab}$ is given
in the double index notation by
\EQ \eta_{(\a \bar \a) (\b \bar\b)} = \frac{1}{2} \bar
\eta_{\a \b} \bar \eta_{\bar \a \bar \b}\ , \ \ {\rm with }\ \
\bar \eta_{12} = \bar \eta_{21} =1,\
 \bar \eta_{11} = \bar \eta_{22} =0.
\EN
Indices are raised and lowered with the tensors $g^{ij},\ \eta_{ab}$
and their inverse.
This notation is basically the same as the one
given in refs.~\cite{STVS}, but we find
that the above double index notation makes the $SU(2) \otimes SU(2)$
group structure more manifest.

The BRST current for the $N=4$ SCA can be derived
directly from the OPEs (2.1) by the method described in
appendix A, and is given by
\bea
J_{BRST}(z)
 &=&  c T + d J + c^\pm_i J^{\pm,i} + \c^a G_a
+ \d^a F_a + b c \pa c + d \pa ( c a) + c^\pm_i \pa ( c b^{\pm, i}) \nn
&+&  \pa c ( {3\over 2} \b_a \c^a + {1\over 2} \a_a \d^a)
+ c (\pa \b_a \c^a + \pa \a_a \d^a)
 - {1\over 2} f^{ij}{}_k c^\pm_i c^\pm_j b^{\pm, k} \nn
&\mp&  2 {k^\pm \over k} R^{\pm,i}{}_a{}^b \pa c^\pm_i \a_b \c^a
+ R^{\pm,i}{}_a{}^b c^\pm_i (\b_b \c^a + \a_b \d^a)
- 4{k^\mp \over k} R^{\pm,i}_{ab} b^\pm_i \pa \c^a \c^b \nn
&\pm&  2 R^{\pm,i}_{ab} b^\pm_i \d^a \c^b
- \c_a \c^a b - \d_a \c^a a + \pa d \a_a \c^a \ ,
\ena
where $(c,b),(d,a),(c_i^\pm,b_i^\pm),(\c^a,\b_a),(\d^a,\a_a)$ are the
reparametrization, $U(1)$ current, $SU(2)$ currents, supersymmetry and
spin-$\shalf$ fermion ghosts, respectively, with correlations
\bea
c(z) b(w) &\sim&\ d(z) a(w) \sim\  { 1\over (z-w)}\ ,\qquad
c^\pm_i(z) b^\pm_j(w) \sim\ {g_{ij}\over (z-w)}\ ,\nn
\c^a(z) \b_b(w) &\sim&\ \d^a(z) \a_b(w) \sim\
{\d^a_b \over (z-w)}\ .
\ena
We will discuss the nilpotency condition for the BRST operator in
sect.~4.
It can be shown that the BRST operator can be cast into the form
\bea
Q &=& \oint\dz \left[ c\left( T + \shalf T_{gh} \right)
+ d \left( J + \shalf J_{gh} \right)  \right. \nn
&+& \left. c_i^A \left( J^{A,i} + \shalf J^{A,i}_{gh} \right)
+ \c^a \left( G_a + \shalf G_{gh,a} \right)
+ \d^a \left( F_a + \shalf F_{gh,a} \right) \right],
\ena
where the generators with subscript $gh$ are those for the ghosts.
The explicit forms of these generators are given in appendix B.

\sect{Large $N=4$ SCA and its BRST operator in $N=2$
superfields}
\indent
The results in the previous section are rather complicated to deal with
if we write all components explicitly.
It is much simpler, in general,
to write everything in terms of $N=2$ superfields, as done in
ref.~\cite{RASS},
even though this makes less manifest the $SU(2) \otimes SU(2)$
group structure.
In this section, we describe the equivalent results in
superfields and make an explicit connection with the components.

Our superspace conventions are as follows:
$ Z \equiv (z,\t,\tb) $
denotes the super-coordinates,
\EQ
D = \pa_\t -\shalf \tb\pa_z, \qquad
{\bar D} = \pa_\tb -\shalf \t\pa_z, \qquad \{ D,\bar D \} = - \pa_z
\EN
are the supercovariant derivatives in the $N=2$ superspace and
\bea
& &z_{12} \equiv z_1 - z_2 +\shalf(\t_1\tb_2 +\tb_1\t_2),\qquad
\t_{12} \equiv \t_1 -\t_2, \qquad
\tb_{12} \equiv \tb_1 -\tb_2, \nn
& &D_1 z_{12} = D_2  z_{12} = -\shalf \tb_{12}, \qquad
\bar D_1 z_{12} = \bar D_2  z_{12} = -\shalf \t_{12} .
\ena
The $N=4$ SCA in $N=2$ superfields is~\cite{RASS}\footnote{
We have absorbed the factor $\sqrt{1-4\a^2}$ into the normalization of
the current $J$ compared with ref.~\cite{RASS}. Here our $x$ corresponds
to their $2\a$.}
\bea
T(Z_1) T(Z_2) &\sim& \frac{\frac{1}{3} c + \t_{12}\tb_{12} T}{z_{12}^2}
  +\frac{-\t_{12}DT + \tb_{12}{\bar D}T + \t_{12}\tb_{12}\pa T}{z_{12}}, \nn
T(Z_1)G(Z_2) &\sim& \shalf\frac{\t_{12}\tb_{12}}{z_{12}^2}G
      + \frac{-\t_{12}DG + \tb_{12}{\bar D}G + \t_{12}\tb_{12}\pa G
      +x G}{z_{12}}, \nn
T(Z_1){\bar G}(Z_2) &\sim& \shalf\frac{\t_{12}\tb_{12}}{z_{12}^2}{\bar G}
      + \frac{-\t_{12}D{\bar G} + \tb_{12}{\bar D}{\bar G} + \t_{12}\tb_{12}
      \pa{\bar G} - x{\bar G}}{z_{12}}, \nn
T(Z_1)J(Z_2) &\sim& \frac{-\t_{12}DJ + \tb_{12}{\bar D}J + \t_{12}\tb_{12}
      \pa J}{z_{12}}, \nn
G(Z_1) {\bar G}(Z_2) &\sim& \frac{kx}{2}\frac{\t_{12}\tb_{12}}{z_{12}^2}
 + \frac{-k-\t_{12}DJ + \tb_{12}{\bar D}J + \t_{12}\tb_{12}
 (-T + \shalf\pa J + \frac{x}{2}[D,{\bar D}]J)}{z_{12}}, \nn
J(Z_1) G(Z_2) &\sim& - \frac{\t_{12}\tb_{12}}{z_{12}} G, \qquad
J(Z_1) {\bar G}(Z_2) \sim  \frac{\t_{12}\tb_{12}}{z_{12}}{\bar G}, \nn
J(Z_1) J(Z_2) &\sim& - 2k \ln z_{12},
\ena
where it is understood that all the operators on the right hand side are
evaluated at the point $Z_2$ and we have defined
\EQ
x \equiv \frac{k^+-k^-}{k},\qquad
c = \frac{3k}{2}(1-x^2).
\EN
The parameter $x$ measures the asymmetry between the two $SU(2)$ current
algebras.

Using the general procedure described in appendix A, we find that
the BRST operator for this algebra is given by
\bea
Q &=& \oint \dZ \left[ C_t T +C_j J + C_g G + C_{\bar g}{\bar G}
 + C_t \left( \shalf \pa C_t B_t
+ \shalf (DC_t)({\bar D}B_t)       \right. \right. \nn
&+&  \shalf ({\bar D}C_t)(DB_t)
- \shalf \pa ( C_g B_g)
+ (DC_g)({\bar D}B_g) + ({\bar D}C_g)(DB_g)
  - \shalf \pa ( C_{\bar g} B_{\bar g} ) \nn
&+& \left.
(DC_{\bar g})({\bar D}B_{\bar g}) + ({\bar D}C_{\bar g})(DB_{\bar g}) +
(DC_j)({\bar D}B_j) + ({\bar D}C_j)(DB_j) \right)
\nn
&+& (C_{\bar g} B_{\bar g} - C_g B_g)\left(C_j + \frac{x}{2}[D,{\bar D}]C_t
\right) + C_g C_{\bar g}\left( B_t -\frac{x}{2}[D,{\bar D}]B_j \right)\nn
&+& \left. C_{\bar g} \left( -\shalf C_g \pa B_j +
(DC_g)({\bar D}B_j) + ({\bar D}C_g)(DB_j) \right)\right],
\ena
where the four sets of fields $(C_t,B_t),(C_g,B_g),(C_{\bar g},B_{\bar g})$
and $(C_j,B_j)$ with spins $(-1,1),(-\shalf,\shalf),$ $(-\shalf,\shalf)$ and
$(0,0)$ are the reparametrization, supersymmetry and current
ghosts in $N=2$ superfields for the large $N=4$ superstrings with the
correlations
\EQ
C(Z_1) B(Z_2) \sim \frac{\t_{12}\tb_{12}}{z_{12}}.
\EN
It can also be shown that the BRST operator (3.5) for this $N=4$
superstring can be written as
\bea
Q &=& \oint\dZ \left[ C_t \left( T + \shalf T_{gh} \right)
      + C_g \left( G + \shalf G_{gh} \right) \right. \nn
&+& \left. C_{\bar g} \left({\bar G} + \shalf{\bar G}_{gh} \right)
      + C_j \left( J + \shalf J_{gh} \right) \right],
\ena
where the generators with subscript $gh$ are those for ghosts.
Their explicit forms are reported in appendix B.
We will examine the nilpotency condition for this BRST charge in the
next section.

The relation between the superfields for generators in (3.3) and
the components in sect. 2 may be worked out. We find that the exact
correspondence is as follows:
\bea
T &=& (1+x)J^{-,3} - (1-x)J^{+,3} + \t G_{(2,1)} +\tb G_{(1,2)}
 + \t\tb T_B \ , \nn
G &=& 2 F_{(1,1)} + \t J^{-,+} + \tb J^{+,+}
 + \t\tb\left(G_{(1,1)}+ x \pa F_{(1,1)}\right)\ ,\nn
{\bar G} &=& 2 F_{(2,2)} + \t J^{+,-} + \tb J^{-,-}
 - \t\tb \left(G_{(2,2)}+ x \pa F_{(2,2)}\right)\ , \nn
J &=& 2 \int dz J_B - 2\t F_{(2,1)} + 2\tb F_{(1,2)}
 - \t\tb(J^{+,3}+J^{-,3})\ ,
\ena
where we have put the subscript $B$ on the component energy-momentum tensor
and $U(1)$ current to avoid confusion.
This result tells us that the components of the ghost fields are
embedded in the ghost superfields as follows
\bea
C_t &=& c + \t \c_{(1,2)} - \tb \c_{(2,1)} + \t\tb \frac{c^-_3
 -c^+_3}{2}\ , \nn
C_g &=& \c_{(1,1)} - \t c^+_- + \tb c^-_-
 + \shalf \t\tb\left(\d_{(1,1)}+x\pa \c_{(1,1)}\right), \nn
C_{\bar g} &=& -\c_{(2,2)} - \t c^-_+ + \tb c^+_+
 + \shalf \t\tb\left(\d_{(2,2)} + x\pa \c_{(2,2)}\right), \nn
2 C_j &=& -(1+x)c^+_3 -(1-x)c^-_3 + \t \d_{(1,2)}
 + \tb \d_{(2,1)} - \t\tb \pa d \ ,
\ena
in order to reproduce the structure for the BRST current
given in eq.~(2.8).
The components of antighosts are then determined from the correlators
(3.6) to be
\bea
B_t &=& (1+x)b^-_3 -(1-x)b^+_3-\t\b_{(2,1)}-\tb\b_{(1,2)} + \t\tb b \ , \nn
B_g &=& 2\a_{(1,1)} - \t b^-_+ - \tb b^+_+
 + \t\tb\left(\b_{(1,1)}+x\pa\a_{(1,1)}\right), \nn
B_{\bar g} &=& 2\a_{(2,2)} - \t b^+_- - \tb b^-_-
 - \t\tb\left(\b_{(2,2)}+x\pa\a_{(2,2)}\right), \nn
B_j &=& 2\int dz a + 2\t\a_{(2,1)}-2\tb\a_{(1,2)}-\t\tb (b^+_3+b^-_3) \ .
\ena
Although $J$ and $B_j$ contain integrals, i.e. $J$ contains the integral
of the bosonic $U(1)$ current $J_B$ and $B_j$ the integral of
the corresponding antighost $a$, this does not cause any
problem because these fields always appear with derivatives in
physical quantities.

A perhaps better way to present the large $N=4$ algebra
in $N=2$ superfields is to use the chiral and antichiral fields
$H\equiv DJ$ and $\bar H \equiv \bar D J$,
instead of the general superfield $J$, as independent fields.
In this way no logarithm will appear in the OPEs and
we will never need to introduce the integral of the bosonic
$U(1)$ current $J_B$. The relevant OPEs in addition to the first
three lines of eq. (3.3) are as follows:
  \bea
T(Z_1) H(Z_2) &\sim& \left( \frac{ \t_{12}\tb_{12} }{z_{12}^2}
+ \frac{2}{z_{12}} \right) \shalf H
  +\frac{ \tb_{12}{\bar D}H + \t_{12}\tb_{12}\pa H}{z_{12}}\ , \nn
T(Z_1) \bar H(Z_2) &\sim& \left( \frac{ \t_{12}\tb_{12} }{z_{12}^2}
- \frac{2}{z_{12}} \right) \shalf \bar H
  +\frac{ - \t_{12} D \bar H + \t_{12}\tb_{12}\pa \bar H}{z_{12}}\ , \nn
G(Z_1) {\bar G}(Z_2) &\sim& \frac{kx}{2}\frac{\t_{12}\tb_{12}}{z_{12}^2}
 + \frac{-k-\t_{12}H + \tb_{12}{\bar H} - \t_{12}\tb_{12}
 (T +  \frac{1 + x}{2}{\bar D}H  +  \frac{1 - x}{2} D \bar H )}{z_{12}}
\ , \nn
H(Z_1)G(Z_2) &\sim& - \frac{\tb_{12}}{z_{12}}G \ , \ \ \ \ \ \ \
H(Z_1)\bar G(Z_2) \sim \frac{\tb_{12}}{z_{12}} \bar G \ ,\nn
\bar H(Z_1)G(Z_2) &\sim&  \frac{\t_{12}}{z_{12}}G \ ,\ \ \ \ \ \ \ \ \
\bar H(Z_1)\bar G(Z_2) \sim - \frac{\t_{12}}{z_{12}}\bar G \ ,\nn
H(Z_1) \bar H(Z_2) &\sim&  - \frac{k}{2} \left(
\frac{ \t_{12}\tb_{12} }{z_{12}^2}
- \frac{2}{z_{12}} \right) .
\ena
The BRST charge can now be written using the chiral
and antichiral ghost superfields
$(C_h,B_h)$ and $(C_{\bar h},B_{\bar h})$ with correlations
\EQ
C_h(Z_1) B_h(Z_2) \sim \frac{\tb_{12}}{z_{12}}, \ \ \ \ \ \
C_{\bar h}(Z_1) B_{\bar h}(Z_2) \sim \frac{\t_{12}}{z_{12}},
\EN
and reads as follows:
\bea
Q &=& \oint \dZ \left[ C_t T  + C_g G + C_{\bar g}{\bar G}
\right]
+ \oint \dZc C_h H +  \oint \dZa C_{\bar h} \bar H \nn
 &+&  \oint \dZ \left[ C_t \left( \shalf \pa C_t B_t
+ \shalf (DC_t)({\bar D}B_t) +\shalf ({\bar D}C_t)(DB_t)
- \shalf \pa ( C_g B_g) \right. \right.\nn
&+& (DC_g)({\bar D}B_g) + ({\bar D}C_g)(DB_g)
- \left. \shalf \pa ( C_{\bar g} B_{\bar g}) +
(DC_{\bar g})({\bar D}B_{\bar g}) + ({\bar D}C_{\bar g})(DB_{\bar g})
\right) \nn
 &+& (\bar D C_h)B_h C_t - (DC_{\bar h})B_{\bar h} C_t
+ (C_{\bar g} B_{\bar g} - C_g B_g)\left(C_{\bar h} - C_h
+\frac{x}{2}[D,\bar D]C_t
\right) \nn
&-&\left. C_{\bar g}\left( (\bar DC_g)B_h + (DC_g)B_{\bar h}\right)
+ C_g C_{\bar g} \left(B_t + \frac{1-x}{2} DB_{\bar h}
-\frac{1+x}{2}\bar DB_h \right)
\right].
\ena

\sect{Nilpotency of the BRST operator}
\indent
So far we have revealed the general structure of the large $N=4$
superconformal theory and constructed the BRST operator. The most important
property of the BRST charge is its nilpotency. In this section, we will
examine what is the necessary and sufficient condition for this
property to hold.

In order for the square of the BRST charge to vanish, the first order pole
of the OPE of the BRST current (2.8) with itself must be zero up to
total derivatives. Actually, we have computed the first order pole of
this OPE and obtained quite a
large number of terms in components. We have checked that they are
indeed total derivatives for the
case of $k^+=k^-$ and $c=0$, which is the $N=4$ supersymmetry realized in
the $N=2$ superstring \cite{GR,BOP}.
It is cumbersome to check the nilpotency of the BRST operator for the
general case by this method.
A much simpler way to check this is to compute the double commutators of
the BRST operator with all the fundamental fields in the theory.
The necessary and sufficient condition for the nilpotency of the charge
is that they all vanish because this means that the square of the BRST
charge vanishes in the Hilbert space of the theory.\footnote{In general,
one can tell if an expression is a total derivative or not by checking
that it has no simple pole in its OPEs with all the fundamental fields.
We find that this method is quite useful.}

We have computed the double commutators and the results are as follows.
The double commutators with the ghosts and with the matter generators
all vanish without any restriction. On the other hand, those with
the antighosts are nonvanishing in general:
\bea
[Q,\{ Q,b\}] &=& \frac{k}{8}(1-x^2) \pa^3 c, \qquad
{}[Q,\{ Q,a\}] = - \frac{k}{2} \pa d, \nn
{}[Q,[Q,\b_a]] &=& \frac{k}{2}(1-x^2) \pa^2 \c_a, \qquad
{}[Q,[Q,\a_a]] = -\frac{k}{4} \d_a, \nn
{}[Q,\{ Q,b^+_i\}] &=&  \frac{k}{4}(1+x) \pa c^+_i, \qquad
{}[Q,\{ Q,b^-_i\}] = \frac{k}{4}(1-x) \pa c^-_i.
\ena
These results show that the necessary and sufficient condition for the
nilpotency of the BRST charge for the large $N=4$ superstring is
$ k=0$ and $x$ arbitrary.
This implies that $c=0$ \cite{KS}, but we see that there is a line of
critical $N=4$ string theories depending on the parameter $x$.
Note also that the converse is not true: $c=0$ allows solutions $x=\pm 1$
with arbitrary $k$.
It is the string theory with $x=0$ that appears in the hierarchy
of ref.~\cite{BOP2}.
We have also reproduced these results using $N=2$ superfields.

\sect{Discussions and conclusions}
\indent
We discuss in this section how to obtain the small $N=4$ SCA
as a subalgebra of the large $N=4$ SCA.
While it is known that the small $N=4$
may be obtained from the large one, the precise
procedure of how to derive it in the $N=2$ superfield notation
has never been given. Here we fill this gap since it may be useful
for recognizing possible embeddings of the small
$N=4$ string into the large one.

The small $N=4$ SCA is a subalgebra of the large $N=4$
when $x=\pm 1$. Let us take the case $x=1$.
By setting $ k = 2 \hat k (1-x)^{-1}$
and taking the limit $x \to 1$, one recognizes from eq. (3.3) the following
small $N=4$ subalgebra
\bea
T(Z_1) T(Z_2) &\sim& \frac{\frac{1}{3}c + \t_{12}\tb_{12} T}{z_{12}^2}
  +\frac{-\t_{12}DT + \tb_{12}{\bar D}T + \t_{12}\tb_{12}\pa T}{z_{12}}, \nn
T(Z_1)G_c(Z_2) &\sim&  \left( \frac{ \t_{12}\tb_{12} }{z_{12}^2}
+ \frac{2}{z_{12}} \right)  G_c
  +\frac{ \tb_{12}{\bar D}G_c + \t_{12}\tb_{12}\pa G_c}{z_{12}}\ , \nn
T(Z_1)G_a(Z_2) &\sim&  \left( \frac{ \t_{12}\tb_{12} }{z_{12}^2}
- \frac{2}{z_{12}} \right)  G_a
  +\frac{- \t_{12} DG_a + \t_{12}\tb_{12}\pa G_a}{z_{12}}\ , \nn
G_c(Z_1) G_a(Z_2) &\sim& - \frac{c}{6}
\left( \frac{ \t_{12}\tb_{12} }{z_{12}^3} - \frac{1}{z_{12}^2} \right)
- \shalf
\left( \frac{ \t_{12}\tb_{12} }{z_{12}^2} - \frac{2}{z_{12}} \right)
T +\frac{ \tb_{12}}{z_{12}} {\bar D}T \ ,
\ena
  where $G_c\equiv DG$ and $G_a \equiv \bar D \bar G $
are chiral and antichiral superfields, respectively,
and $c= 6 \hat k$. The parameter $\hat k$ is the level of the
remaining $SU(2)$ current algebra contained in the small $N=4$ SCA.
One can proceed similarly for the case $x=-1$.
The BRST charge in the $N=2$ superfield notation can also be
easily constructed.
Using the ghosts with correlators
\EQ
C_t(Z_1) B_t(Z_2) \sim \frac{\t_{12}\tb_{12}}{z_{12}}, \ \ \ \ \
C_c(Z_1) B_c(Z_2) \sim \frac{\tb_{12}}{z_{12}}, \ \ \ \ \ \
C_a(Z_1) B_a(Z_2) \sim \frac{\t_{12}}{z_{12}},
\EN
where $(C_t,B_t)$, $(C_c, B_c)$ and $(C_a,B_a)$ are general,
chiral and antichiral
superfields, respectively, one can write the BRST charge as follows
\bea
Q &=& \oint \dZ  C_t T  + \oint \dZc C_c G_c +  \oint \dZa C_a G_a \nn
 &+&  \oint \dZ \left[ C_t \left( \shalf \pa C_t B_t
+ \shalf DC_t{\bar D}B_t +\shalf {\bar D}C_t DB_t  \right. \right.\nn
&+&\left. \left.
2 DC_a B_a +  C_a DB_a - 2 {\bar D} C_c B_c - C_c {\bar D} B_c \right)
- B_t C_c C_a \right].
\ena
We checked that it is nilpotent for $c=-12$.

For completeness, we review also how the $N=3$ SCA is contained into
the large $N=4$ SCA. The $N=3$ subalgebra is recognized for $x=0$
by defining $\hat G = G - \bar G$, and reads
\bea
T(Z_1) T(Z_2) &\sim& \frac{\frac{1}{3} c + \t_{12}\tb_{12} T}{z_{12}^2}
  +\frac{-\t_{12}DT + \tb_{12}{\bar D}T + \t_{12}\tb_{12}\pa T}{z_{12}}, \nn
T(Z_1)\hat G(Z_2) &\sim& \shalf\frac{\t_{12}\tb_{12}}{z_{12}^2} \hat G
      + \frac{-\t_{12}D \hat G + \tb_{12}{\bar D}\hat G +
\t_{12}\tb_{12}\pa \hat G }{z_{12}}, \nn
\hat G(Z_1) \hat G(Z_2) &\sim& \frac{\frac{4}{3} c
+ 2 \t_{12} \tb_{12} T}{z_{12}} \ .
\ena
It is clear from this derivation that the small $N=4$ SCA
does not contain the $N=3$ SCA as its subalgebra.
Obviously both contain the $N=2$ SCA whose OPE is given
by the first line of (5.1).

To conclude, we recall that we have constructed the BRST operator
for the large $N=4$ SCA and showed that its nilpotency requires the
vanishing of the usual Virasoro central charge, but it is consistent
with the existence of a one-parameter family of string theories
labeled by a parameter $x$. It would be interesting to explicitly
construct and analyze some of these string theories, even though they
do not seem to possess an interesting space-time interpretation.
As far as the search for an \lq\lq universal\rq\rq \  string theory
started by Berkovits and Vafa,
it seems possible that the small $N=4$
strings can also be obtained from the large $N=4$ ones at the values
$x=\pm1$, and that the large $N=4$ string theories at arbitrary
values of $x$ may be embedded into the $N=5$ string.
If this can be achieved one has the result that all string theories
based on the linear SCAs belong to a chain of
embeddings, thus identifying  the $N=\infty$ string theory as
the master theory which generates all the other ones by successive
symmetry breaking.

\vs{10}
\noindent{\em Acknowledgements}

We would like to thank J. L. Petersen for useful discussions
and N. Berkovits for valuable comments and suggestions.
One of us (N. O.) would like to thank P. Di Vecchia for discussions,
support and kind hospitality at NORDITA where this work was carried out.
Many of the calculations in this paper were done by using the OPE
packages developed by K. Thielemans
and S. Krivonos, whose software is gratefully acknowledged.

\newpage
\appendix
\noindent
{\Large\bf Appendix}

\sect{General method for constructing the BRST operator}
\indent
In this appendix, we describe how to obtain the BRST operator directly
from the operator algebra.
Let us start from a graded Lie algebra. Suppose we are given
bosonic generators $T^a$ and fermionic generators $G^\a$
satisfying the algebra
\bea
[T^a,T^b] &=& f^{ab}{}_c T^c \ , \nn
{}[T^a,G^\a] &=& f^{a\a}{}_\b G^\b \ , \nn
\{ G^\a,G^\b\} &=& f^{\a\b}{}_a T^a \ .
\ena
To construct the BRST operator we introduce ghosts and antighosts,
i.e. anticommuting fields $(c_a, b^a)$ for the bosonic generators
$T^a$ and commuting fields $(\c_\a , \b^\a)$
for the fermionic generators $G^\a$. They have the following
(anti)commutation relations
\EQ
\{ c_a,b^b\} = \d_a^b \ , \qquad
[\c_\a,\b^\b] = \d_\a^\b \ .
\EN
Using the Jacobi identities,
it is easy to show that the BRST operator given by
\EQ
Q = c_a T^a + \c_\a G^\a + \shalf f^{ab}{}_c b^c c_b c_a
 + f^{a\a}{}_\b \b^\b \c_\a c_a - \shalf f^{\a\b}{}_a b^a \c_\b \c_\a \ ,
\EN
is nilpotent.
This rule can be rewritten as follows. Take the terms containing the
structure constants in eq. (A.3). They may be written as
\EQ
-\shalf \frac{\pa_r}{\pa\Lambda} c_a [T^a,T^b]c_b
+ \frac{\pa_r}{\pa\Lambda} c_a [T^a,G^\a]\c_\a
-\shalf \frac{\pa_r}{\pa\Lambda} \c_\a \{G^\a,G^\b\}\c_\b \ ,
\EN
where it is understood that the generators appearing in the (anti)commutators
are replaced by $b^c\Lambda$ for $T^c$ and $\b^\c\Lambda$ for $G^\c$,
with an anticommuting parameter $\Lambda$ removed from the right
by $ \frac{\pa_r}{\pa \Lambda}$.

This result can be used to derive the BRST operator
directly from an operator algebra. Suppose we are given the OPEs among
$T^a(z)$ and $G^\a(z)$.
We introduce ghost and antighost fields with correlators
\EQ
c_a(z) b^b(w) \sim \frac{\d_a^b}{(z-w)} \ , \ \ \ \
\c_\a(z) \b^\b (w) \sim \frac{\d_\a^\b}{(z-w)} \ .
\EN
The leading term in the BRST charge is then given by
\EQ
\oint ( c_a T^a + \c_\a G^\a ) \ ,
\EN
while the terms corresponding to the structure constants are obtained
form the above prescription and read
\bea
&&-\shalf \frac{\pa_r}{\pa\Lambda} \oint_w\oint_z c_a(z) T^a(z)T^b(w)c_b(w)
+ \frac{\pa_r}{\pa\Lambda} \oint_w\oint_z c_a(z) T^a(z)G^\a(w)\c_\a(w) \nn
&&-\shalf\frac{\pa_r}{\pa\Lambda} \oint_w\oint_z
 \c_\a(z)G^\a(z)G^\b(w)\c_\b(w) \ ,
\ena
where it is understood that the products of operators are replaced by
their OPEs, and those generators appearing in the OPEs are replaced by
the products of antighosts and $\Lambda$.

As a simple example, let us consider the Virasoro algebra,
\EQ
T(z)T(w) \sim \frac{2T(w)}{(z-w)^2} + \frac{\pa T(w)}{(z-w)} \ .
\EN
We get
\EQ
-\shalf \frac{\pa_r}{\pa\Lambda} \oint_w\oint_z c(z)\left[
 \frac{2b(w)\Lambda}{(z-w)^2} + \frac{\pa
b(w)\Lambda}{(z-w)}\right]c(w) \ ,
\EN
which after the $z$ integration gives correctly
\EQ
\oint \pa c b c \ .
\EN

It is easy to show that the same prescription is valid for
the $N=2$ superfields if we write everything in terms of
superfields and replace integrals by superspace integrals.
It would be interesting to investigate if this method could be
extended to the case of nonlinear algebras~\cite{SSPN}.

\sect{Ghost generators}
\indent
In this appendix, we summarize all the ghost generators. In components
we have
\bea
T_{gh} &=& -2 b \pa c - \pa b c - a \pa d - b^{\pm i} \pa c_i^\pm
- \frac{3}{2}\beta_a \pa\gamma^a - \shalf \pa \beta_a \gamma^a
- \shalf \a_a \pa \d^a + \shalf \pa \a_a \d^a \ , \nn
J_{gh} &=& \pa(c a - \a_a \c^a) \ ,\nn
J^{+,i}_{gh} &=& \pa(c b^{+, i})  -  f^{ij}{}_k  c^+_j b^{+, k}
+  2 {k^+ \over k} R^{+,i}{}_a{}^b \pa (\a_b \c^a )
+ R^{+,i}{}_a{}^b  (\b_b \c^a + \a_b \d^a) \ , \nn
J^{-,i}_{gh} &=& \pa(c b^{-, i})  -  f^{ij}{}_k  c^-_j b^{-, k}
-  2 {k^- \over k} R^{-,i}{}_a{}^b \pa (\a_b \c^a )
+ R^{-,i}{}_a{}^b  (\b_b \c^a + \a_b \d^a) \ , \nn
G_{gh,a} &=&   {3\over 2} \pa c \b_a + c \pa \b_a
\mp  2 {k^\pm \over k} R^{\pm,i}{}_a{}^b \pa c^\pm_i \a_b
+ R^{\pm,i}{}_a{}^b c^\pm_i \b_b
+ 4{k^\mp \over k} R^{\pm,i}_{ab} \pa b^\pm_i  \c^b \nn
&& + 8{k^\mp \over k} R^{\pm,i}_{ab}  b^\pm_i  \pa \c^b
\pm 2 R^{\pm,i}_{ba} b^\pm_i \d^b
-2 \c_a  b - \d_a  a + \pa d \a_a \ , \nn
F_{gh,a} &=& {1\over 2}\pa c \a_a + c  \pa \a_a
+ R^{\pm,i}{}_a{}^b c^\pm_i \a_b
\pm 2 R^{\pm,i}_{ab} b^\pm_i \c^b
-\c_a a \ .
\ena
It is interesting to note that the $U(1)$ current $J_{gh}$
is a total derivative.
It is also easy to check that these generators
satisfy the OPE (2.1) with $c=0$.

The ghost generators in $N=2$ superfields are as follows
\bea
T_{gh} &=& - \pa(B_t C_t) + ({\bar D}C_t)(DB_t) + (DC_t)({\bar D}B_t)
+ ({\bar D}C_j)(DB_j) + (DC_j)({\bar D}B_j) \nn
&-& \shalf\pa(B_g C_g) +(1+x)({\bar D}C_g)(DB_g)+(1-x)(DC_g)({\bar D}B_g) \nn
&-& \shalf\pa(B_{\bar g} C_{\bar g}) + (1-x)({\bar D}C_{\bar g})(DB_{\bar g})
 +(1+x)(DC_{\bar g})({\bar D}B_{\bar g}) \nn
&-& \frac{x}{2}\left( [D,{\bar D}]C_g B_g + C_g[D,{\bar D}]B_g\right)
 + \frac{x}{2}\left( [D,{\bar D}]C_{\bar g}B_{\bar g}
 + C_{\bar g} [D,{\bar D}] B_{\bar g}\right) , \nn
G_{gh} &=& \shalf C_{\bar g} \pa B_j + C_t \pa B_g + \shalf \pa C_t B_g
 + C_{\bar g}B_t + ({\bar D}C_t)(DB_g) + (DC_t)({\bar D}B_g) \nn
&-& C_j B_g - ({\bar D}C_{\bar g})(DB_j)
  - (DC_{\bar g})({\bar D}B_j) - \frac{x}{2} \left( B_g[D,{\bar D}]C_t
 + C_{\bar g}[D,{\bar D}]B_j\right), \nn
{\bar G}_{gh} &=& -\shalf C_g \pa B_j + C_t \pa B_{\bar g}
 + \shalf \pa C_t B_{\bar g} + C_g B_t
 + (DC_t)({\bar D}B_{\bar g}) + ({\bar D}C_t)(DB_{\bar g}) \nn
&+& C_j B_{\bar g} + (DC_g)({\bar D}B_j) + ({\bar D}C_g)(DB_j)
 + \frac{x}{2}\left( B_{\bar g}[D,{\bar D}]C_t
 - C_g[D,{\bar D}]B_j \right), \nn
J_{gh} &=& C_t \pa B_j - C_g B_g + C_{\bar g}B_{\bar g}
 + (DC_t)({\bar D} B_j) + ({\bar D}C_t)(DB_j).
\ena
It is easy to check that these generators  satisfy the OPE (3.3)
with $c=k=0$.

\end{document}